\documentclass[11pt,twoside]{article}
\usepackage{asp2004}
\usepackage{psfig}
\usepackage{epsf}
\usepackage{graphics}
\usepackage{lscape}
\def\edcomment#1{\iffalse\marginpar{\raggedright\sl#1\/}\else\relax\fi}
\reversemarginpar

\begin{document}
\title{
AU Scale Structures in Extra-planar Gas
}
\author{P. Richter}
\affil{
Institut f\"ur Astrophysik und Extraterrestrische Forschung, \\ Universit\"at Bonn,
Auf dem H\"ugel 71, 53121 Bonn, Germany
}

\begin{abstract}

Recent spectroscopic observations of intermediate- and high-velocity
clouds (IVCs and HVCs) in the Milky Way halo
have unveiled the presence of diffuse interstellar molecular hydrogen 
(H$_2$) several kpc away from the Galactic disk. Most of this H$_2$ gas
appears to reside in relatively small ($\sim 0.1$ pc), dense
($n_{\rm H}\approx 30$ cm$^{-3}$) gaseous filaments that probably
are part of the cold neutral medium (CNM) in IVCs and HVCs.
Also much smaller structures at AU scale and
very high densities ($n_{\rm H}\approx 800$ cm$^{-3}$) have
been observed, suggesting the presence of tiny-scale 
atomic structures (TSAS) in the Milky Way's 
extra-planar gas. It is not yet understood how such
objects can form and exist in the Milky Way halo, but
the high detection rate of H$_2$ absorption in IVCs 
implies that the CNM represents a gas phase that
is characteristic for neutral clouds in the lower halo.  

\end{abstract}

\thispagestyle{plain}

\section{Introduction}

Significant progress has been made over the last few years
to understand the distribution and origin of extra-planar gas
in the halo of the Milky Way. Studies of the
metal content of intermediate- and high-velocity
clouds - neutral gas clouds embedded in the hot halo with radial 
velocities different from those expected from galactic rotation -
have shown that various different processes contribute
to the neutral gas flow in the halo. Intermediate-velocity
clouds (IVCs) appear to located closer to the disk at
$z$-heights $<3$ kpc, typically. They have metal abundances
similar to those found in the disk of the Milky Way 
(Richter et al.\,2001a, 2001b) and 
probably represent gas that is circulating from the 
Milky Way disk into the halo and back, for instance as
part of a Galactic Fountain (Shapiro \& Field 1976). In contrast, most
of the high-velocity clouds (HVCs) appear to have metallicities
lower than in the Milky Way disk (Wakker et al.\,1999; Lu et al.\,1998;
Richter et al.\,2001b) and thus are probably extragalactic 
in origin. These measurements demonstrate that 
the Milky Way is still in a formation process, accreting
gas and stars from satellite galaxies and the intergalactic
medium, and expelling gaseous material into the halo
as a result of star formation in the disk.

Next to the distribution and origin of the gas in the Milky Way
halo, its physical properties are of great interest as well.
The Galactic halo gas is an extreme multi-phase medium
with temperatures ranging from $50$ to several million
degrees Kelvin. Although the large HVC complexes like
Complex C and the Magellanic Stream span several kpc
in the halo, small-scale structure in this gas is present 
down to scales of several AU (e.g., Meyer \& Lauroesch 1999).
Without beeing influenced by local star formation, the 
the Milky Way's extra-planar gas
represents an excellent laboratory to study 
physcial processes in the diffuse interstellar and
intergalactic medium.

\section{H$_2$ as diagnostics for small-scale structure in the halo}

Molecular hydrogen is 
an excellent diagnostic tool to investigate physical conditions
in the interstellar and intergalactic medium (ISM and IGM, respectively). 
A large number of H$_2$ absorption lines from the Lyman and Werner band
are available in the far-ultraviolet (FUV) in the range 
between $900$ and $1200$ \AA. The {\it Orbiting and Retrievable 
Far and Extreme Ultraviolet Spectrometer} (ORFEUS) was the 
first instrument that allowed us to study H$_2$ absorption in IVCs and
HVCs, but these ORFEUS observations (due to the low sensitivity) 
were limited to only a few 
stellar background sources in the halo and the Magellanic Clouds (Richter 
et al.\,1999; Gringel et al.\,2000; Bluhm et al.\,2001).
With the availability of the {\it Far Ultraviolet Spectroscopic Explorer}
(FUSE) in 1999 (e.g., Moos et al.\,2000) it has become possible 
to systematically study the distribution and abundance of diffuse H$_2$ 
in the halo along a large number of sight lines towards 
stars in the halo and the Magellanic Clouds, and towards quasars 
(e.g., Richter et al.\,2001;
Sembach et al.\,2001; Richter et al.\,2003c). 

The H$_2$ abundance in the diffuse ISM is 
balanced by the formation
of molecules on the surface of dust grains and the H$_2$ destruction 
by the dissociating UV radiation. The volume densities
of H$_2$ and H\,{\sc i}, $n$(H\,{\sc i}) and $n$(H$_2$), 
are linked to the total hydrogen volume density, $n_{\rm H}$,
the H$_2$ grain formation rate, $R$, and the photoabsorption rate, 
$\beta_0$, in a formation-dissociation equilibrium (Spitzer 1978;
Richter et al.\,2003):

\begin{equation}
\frac{n({\rm H\,I})}{n({\rm H}_2)} =
\phi \, \frac{N({\rm H\,I})}{N({\rm H}_2)} =
\frac{\langle k \rangle \beta_0}{Rn_{\rm H}}.
\end{equation}

In this equation, $\langle k \rangle \approx 0.11$ is the
probability that the molecule is dissociated after photoabsorption,
$N$(H\,{\sc i}) and $N$(H$_2$) are the measured H\,{\sc i} and H$_2$
column densities, and $\phi\leq1$ is a scaling factor that accounts
for the possibility that only a fraction of the H\,{\sc i} is physically
related to the H$_2$ gas.
For known photoabsorption and grain formation rates and $\phi$ one
can use equation (1) to estimate gas volume densities 
from measured H\,{\sc i} and H$_2$ column densities. The 
size of an H$_2$ absorbing structure, $D$, then can easily be calculated
from $N$(H\,{\sc i} and $N_{\rm H}$
via $D=\phi\,N$(H\,{\sc i})\,$n_{\rm H}\,^{-1}$.

Also the rotational excitation of the H$_2$ molecules can be 
used to investigate phyiscal properties of the ISM and IGM.
The lowest rotational energy states of H$_2$ (rotational
levels $J=0$ and $1$) are usually excited by collisions, so
that the column density ratio $N(1)/N(0)$ serves as a measure
for the kinetic temperature of the gas. If one plots the 
column densities $N(1)$ and $N(2)$ (divided by the 
quantum-mechanical statistical weight, $g_J$) against the
excitation energy, $E_J$, one can derive the temperature
$T_{01}$ by fitting a Boltzmann distribution to the 
data points. Also higher rotational states (e.g., $J=3,4$ and $5$) 
normally are excited, but for these states other excitation mechanisms such
as UV photon pumping and H$_2$ formation pumping are often dominating.

\section{Detections of H$_2$ absorption in IVCs and HVCs}

A number of positive detections of H$_2$ absorption with FUSE and
ORFEUS have been reported for both IVCs and HVCs, as listed in 
Table 1. The first detection of H$_2$ in a Galactic halo cloud 
was presented by Richter et al.\,(1999) in the high-velocity gas in front
of the Large Magellanic Cloud. Since then, H$_2$ absorption in
halo gas has been found in several intermediate- and high-velocity
clouds, such as the IV Arch, LLIV Arch, IV Spur, Complex gp, 
Draco cloud, LMC-IVC, LMC-HVC, and the Magellanic Stream 
(see Table 1 for details). In all cases, the observed column
densities are low (log $N$(H$_2)\leq 17$), implying
that the H$_2$ resides in a predominantly neutral gas phase.
As expected, H$_2$ absorption preferentially occurs in 
halo clouds that have a high metal and dust abundance
(see, e.g., Richter et al.\,1999)
As an example, we show in Fig.\,1 the FUSE spectrum of the quasar
PG\,1351+640 in the range between $1076.5$ and $1079.5$ \AA,
where a number of H$_2$ lines from various rotational states
are present. Halo H$_2$ absorption at negative intermediate velocities 
from gas in the IV Arch (core IV\,19)
is clearly visible in the various H$_2$ lines shown.
The two-component H$_2$ absorption pattern (Milky Way disk
absorption near zero velocities, IVC H$_2$ absorption near
$-50$ km\,s$^{-1}$) can be well approximated by a two-component 
Gaussian fit (solid line).

\begin{figure}[t!]
\plotone{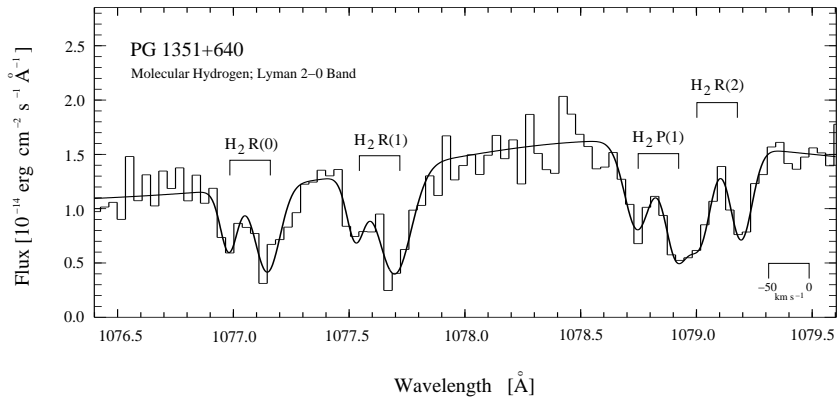}
\caption{FUSE spectrum of the quasar
PG\,1351+640 in the range between $1076.5$ and $1079.5$ \AA.
Next to local disk absorption near zero velocities, 
H$_2$ absorption near $-50$ km\,s$^{-1}$ related to gas
of the Intermediate-Velocity Arch in the halo is clearly
visible in the various H$_2$ lines shown. A two-component 
Gaussian fit (solid line) is overlaid.
}
\end{figure}

\begin{table}[ht!]
\caption{Detections of H$_2$ absorption in Galactic IVCs and HVCs}
\begin{scriptsize}
\begin{center}
\begin{tabular}{lcclclrl}
\noalign{\smallskip}
\tableline
\noalign{\smallskip}
Target & $l$   & $b$   & Cloud Name & $v_{\rm LSR}$ & log $N$(H$_2)$ & log $f^a$ & Ref.$^c$\\
       & (deg) & (deg) &            & [km\,s$^{-1}$] \\
\noalign{\smallskip}
\tableline
\noalign{\smallskip}
\multicolumn{8}{c}{Intermediate-Velocity Clouds}\\
\noalign{\smallskip}
\tableline
\noalign{\smallskip}
       Mrk\,509     & 36.0  & $-$29.9 & Complex gp             &  $+60$ & 14.9$\pm$0.5 & $-$4.3 & 1\\
       Mrk\,876     & 98.3  & $+$40.4 & Draco                  &  $-30$ & 15.6$\pm$0.2 & $-$4.0 & 1\\
       Mrk\,59      & 111.5 & $+$82.1 & IV Arch                &  $-44$ & 14.7$\pm$0.2 & $-$4.3 & 1\\
       PG\,1351+640 & 111.9 & $+$52.0 & IV Arch (IV\,16)       &  $-47$ & 16.4$\pm$0.1 & $-$3.3 & 1\\
       HD\,121800   & 113.0 & $+$49.8 & IV Arch                &  $-70$ & 14.3$\pm$0.6 & $-$5.3 & 1\\
       PG\,1259+593 & 120.6 & $+$58.1 & IV Arch                &  $-54$ & 14.1$\pm$0.2 & $-$5.1 & 2\\
       PG\,0804+761 & 138.3 & $+$31.0 & LLIV Arch              &  $-55$ & 14.7$\pm$0.3 & $-$4.5 & 3\\
       PG\,0832+675 & 147.8 & $+$35.1 & LLIV Arch              &  $-50$ & 15.8$\pm$0.3 & $-$3.9 & 1\\
       NGC\,4151    & 155.1 & $+$75.1 & IV Arch (IV\,26)       &  $-29$ & 15.4$\pm$0.1 & $-$4.5 & 1\\
       NGC\,3310    & 156.6 & $+$54.1 & IV Arch                &  $-47$ & 15.0$\pm$0.8 & $-$4.5 & 1\\
       HD\,93521    & 183.1 & $+$62.2 & IV Arch                &  $-62$ & 14.6$\pm$0.4 & $-$4.7 & 4\\
       PG\,1116+215 & 223.4 & $+$68.2 & IV Spur                &  $-42$ & 15.3$\pm$0.3 & $-$4.3 & 1\\
       HD\,100340   & 258.9 & $+$61.2 & IV Spur                &  $-29$ & 16.0$\pm$0.8 & $-$3.7 & 1\\
       Sk\,-68\,82  & 279.3 & $-$32.8 & IVC toward LMC         &  $+55$ & present$^b$  & ...$^b$ & 5,6,7\\
       Sk\,-60\,80  & 279.3 & $-$32.8 & IVC toward LMC         &  $+50$ & 14.6$\pm$0.5 & $-$1.2 & 6\\
       3C\,273      & 290.0 & $+$64.4 & ...                    &  $+25$ & 15.7$\pm$0.2 & $-$3.4 & 1\\
\noalign{\smallskip}
\tableline
\noalign{\smallskip}
       \multicolumn{8}{c}{High-Velocity Clouds}\\
\noalign{\smallskip}
\tableline
\noalign{\smallskip}
       Sk\,-68\,82  & 279.3 & $-$32.8 & HVC toward LMC         & $+$120 & present$^b$ & ...$^b$ & 5,6,7\\
       NGC\,3783    & 287.5 & $+$23.0 & Lead.\,Arm of the MS  & $+$240 & 16.8$\pm$0.1 & $-2.9$ & 8\\
       Fairall 9    & 295.1 & $-$57.8 & Magellanic Stream      & $+$190 & 16.4$^{+0.3}_{-0.5}$ & $-$3.3 & 2\\
\noalign{\smallskip}
\tableline
\noalign{\smallskip}
\end{tabular}
\end{center}
\noindent $^a$ $f$=2$N$(H$_2$)/[$N$(H\,{\sc i})$+2N$(H$_2$)]\\
\noindent $^b$ H$_2$ is detected, but the H$_2$ column density is highly uncertain\\
\noindent $^c$ References:
       1) Richter et al.\ 2003;
       2) Richter et al.\ 2001c;
       3) Richter et al.\ 2001a;
       4) Gringel et al.\ 2000;
       5) Bluhm et al.\ 2001;
       6) Richter, Sembach \& Howk 2003;
       7) Richter et al.\ 1999;
       8) Sembach et al.\ 2001
\end{scriptsize}
\end{table}

\section{H$_2$ and the cold-neutral medium in the halo}

With FUSE, we have systematically studied 
the properties of the H$_2$ gas in IVCs towards a large number
(56) of mostly extragalactic background sources 
(Richter et al.\,2003). The sample includes 61 IVC
components with H\,{\sc i} column densities 
$\geq 10^{19}$
cm$^{-2}$ and radial velocities 
$25 \leq |v_{\rm LSR}| \leq 100$ km\,s$^{-1}$.
In FUSE spectra with good signal-to-noise
ratios (S/N$>8$ per resolution element) we found 14 clear
detections of H$_2$ in IVC gas with H$_2$ column densities
between $10^{14}$ and $10^{17}$ cm$^{-2}$ (see also Table 1).
In lower S/N data, H$_2$ absorption in IVC gas was tentatively 
detected in additional 17 cases. The molecular hydrogen fraction in 
these clouds, $f=2N$(H$_2)/[(N$(H\,{\sc i})$+2N($H$_2)]$, varies
between $10^{-6}$ and $10^{-3}$. This suggests that the H$_2$  
lives in a relatively dense, mostly neutral gas phase that
probably is linked to the cold neutral medium (CNM) in these clouds.
We now can use equation (1) to determine the
hydrogen volume density and the thickness of the absorbing structure.
The H$_2$ photoabsorption rate in the halo, $\beta_0$, 
depends on the mean ultraviolet radiation field at a height
$z$ above the Galactic plane. The models of Wolfire et al.\,(1995)
predict that that the radiation field at $\sim 1$ kpc above
the disk is approximately 50 percent of that within the disk,
suggesting that the photoabsorption rate in the halo
is $\sim 2.5 \times 10^{-10}$ s$^{-1}$ (see Richter et al.\,2003).
If we now assume that the H$_2$ grain formation rate in
IVCs, $R$, is roughly similar to that within the disk, and 
further set $\phi=0.5$, the
H$_2$ and H\,{\sc i} column densities measured for our IVC
sample imply mean H\,{\sc i} volume densities of $n_{\rm H}
\approx 30$ cm$^{-3}$ and linear diameters of the H$_2$ 
absorbing structures of $D\approx 0.1$ pc. Moreover, if
one considers the rotational excitation of the halo H$_2$ gas 
that can be measured for some of the IVC sight lines, 
one finds for the kinetic temperature of this gas a conservative upper limit
of $T_{\rm kin} \leq 300$ K. Given the relatively high detection
rate of H$_2$ in these clouds, the measurements indicate that 
the CNM phase in IVCs is ubiquitous. Most likely,
the CNM filaments are embedded in a more tenuous gas phase
that corresponds to the warm neutral medium (WNM).

\section{H$_2$ and AU-scale atomic structures in the halo}

\begin{figure}[t!]
\plotone{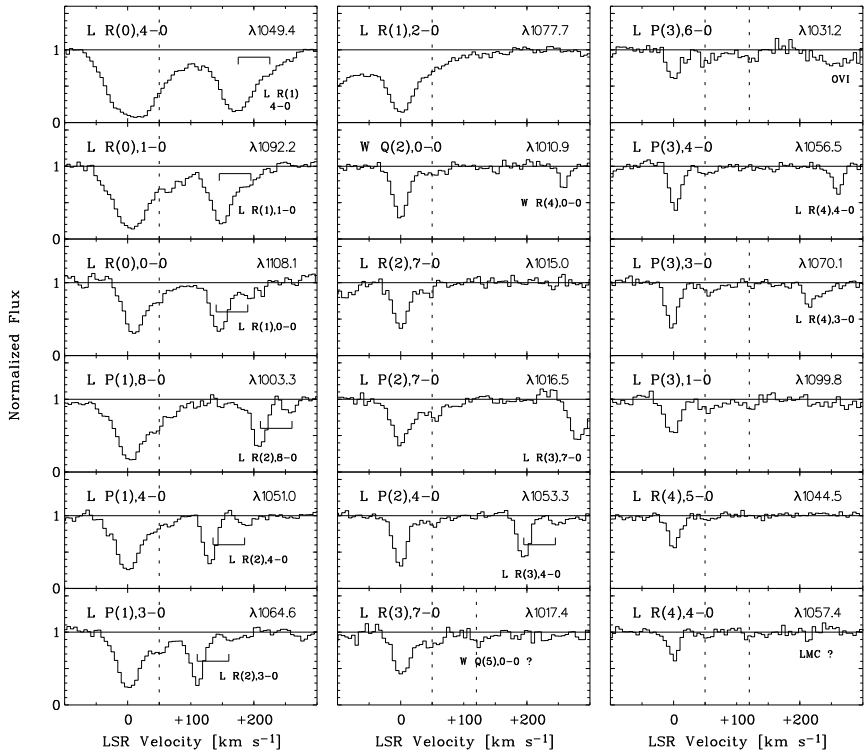}
\caption{Selection of H$_2$ absorption profiles towards the LMC star
Sk $-$68\,80 (Richter, Sembach, \& Howk 2003). H$_2$ halo absorption is detected
at intermediate velocities near $+50$ km\,s$^{-1}$ and possibly also
at high velocities near $+120$ km\,s$^{-1}$ (dashed lines).}
\end{figure}

\begin{figure}[t!]
\plotone{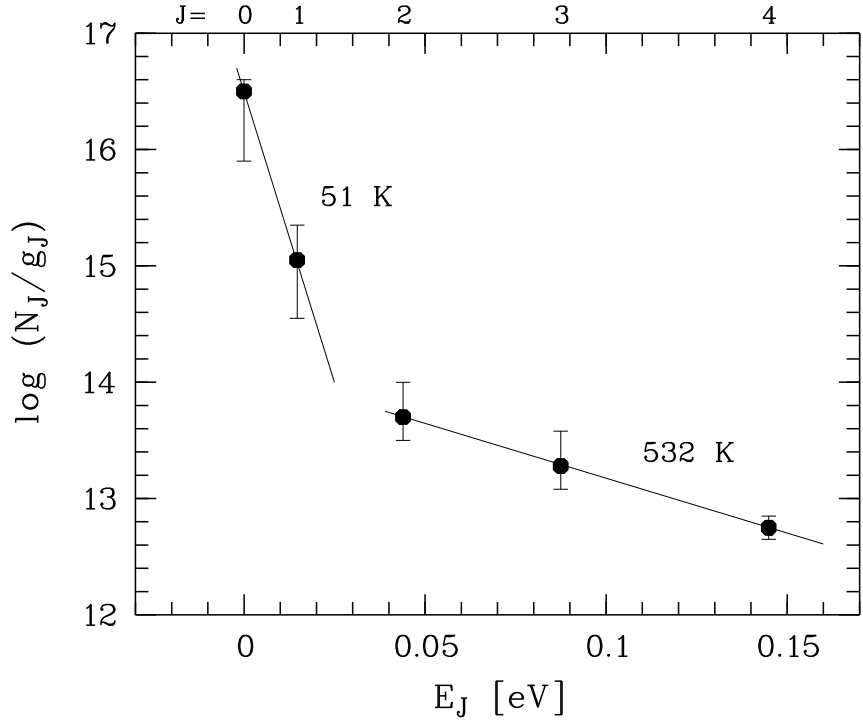}
\caption{Rotational excitation of the IVC H$_2$ gas towards Sk $-$68\,80.
The two rotational ground states ($J=0,1$) can be fitted to an equivalent
Boltzmann temperature of $T=51 \pm 11$ K, providing a measure for 
the kinetic temperature of the H$_2$ absorbing gas.}
\end{figure}

A particularly interesting region to study small-scale
structure in the Milky Way halo is the intermediate- and
high-velocity gas in direction of the Large Magellanic 
Cloud (LMC). 
In this gas complex, H$_2$ absorption in
intermediate- and high-velocity halo gas was detected
for the first time based on data obtained with the
ORFEUS instrument (Richter et al.\,1999).
We recently have reanalyzed the H$_2$ content of the IVC 
towards the LMC using high S/N FUSE data of LMC
background stars (Richter, Sembach, \& Howk 2003). 
Towards the bright LMC star Sk $-$68\,80
H$_2$ absorption at intermediate velocities near 
$+50$ km\,s$^{-1}$ (LSR) is particularly well
defined. Weak H$_2$ absorption in the HVC near 
$+120$ km\,s$^{-1}$ is detected, too.
Also other sight lines towards LMC stars possibly exhibit H$_2$ 
absorption at intermediate and high velocities, but their
spectra often are difficult to analyze because of blending
problems and irregular stellar background continua. 
Towards Sk $-$68\,80 we have detected H$_2$ absorption
in the $+50$ km\,s$^{-1}$ IVC gas in 30 lines from rotational
states $J=0$ to $4$. Some examples for the H$_2$ absorption profiles 
are shown in Fig.\,2. The total H$_2$ column density in the IVC is
log $N$(H$_2)=16.6 \pm 0.5$ together with a very low
$b$ value of $1.5^{+0.8}_{-0.2}$ km\,s$^{-1}$. The low
$b$ value implies that the H$_2$ absorbing gas resides
in a very confined region with a low velocity dispersion. 

The presence of H$_2$ in this cloud is
surprising, given the fact that the measured O\,{\sc i} column
density implies a rather low neutral hydrogen column density
of $N$(H\,{\sc i}$\sim 10^{18}$ cm$^{-2}$. This value is consistent with
the upper limit of $\sim 2 \times 10^{18}$ cm$^{-2}$ from
Parkes 21cm emission observations, although beam smearing 
effects may apply.
Despite the fact that the H\,{\sc i} column density is low, the fraction of
hydrogen in molecular form, $f$, is as high as $\sim 0.07$.
Such a relatively high molecular fraction at this low
total gas column density is unusual considering previous
measurements of H$_2$ absorption in the Milky Way disk (e.g.,
Savage et al.\,1977). If the H$_2$ gas stays in a 
formation-dissociation equilibrium (equation (1)) and 
if H$_2$ self-shielding applies, the measured values for 
$N$(H\,{\sc i}) and $N$(H$_2$) suggest that the molecular
gas resides in a small, dense gaseous filament at a volume 
density of $n_{\rm H} \approx 800$ cm$^{-3}$ and a
linear diameter of only $6.2 \times 10^{14}$ cm or 
$\sim 41$ AU. Possibly, this filament corresponds
to a tiny-scale atomic structure (TSAS; Heiles 1997). Such
AU scale gaseous structures have been 
found in the Milky Way disk 
using H\,{\sc i} 21cm absorption line measurements
(e.g., Faison et al.\,1998). The analysis of the 
H$_2$ rotational excitation for $J=0$ and $1$ 
yields a kinetic gas temperature of only 
$T=51 \pm 11$ K (see Fig.\,3). This temperature is lower than
what is found on average in local H$_2$ gas in the disk 
($\sim 80$ K; Savage et al.\,1977). The thermal pressure
in this cloud then can be derived from our estimates of
$n_{\rm H}$ and $T$ and comes out to be $\sim 5 \times 10^4$ 
cm$^{-3}$ K, about $13$ times higher than the standard thermal
pressure in the CNM in the disk.
It is not yet clear how TSAS with such overpressures 
can form and survive in the otherwise very diffuse Galactic halo gas,
and more data is required to investigate the physical properties
of these objects in more detail.
Unfortunately, the line of sight towards Sk $-$68\,80 so far
remains the only one that shows an H$_2$ absorbing structure in the 
the halo with such extreme properties. The lack of 
further H$_2$ detections of TSAS in the halo is not surprising, however.
In view of their extremely small size, such structures should
have a very low area filling factor. Thus, the chance to
find a TSAS in front of a UV background source is very low.

\section{Conclusions}

Our study of diffuse molecular hydrogen absorption in Galactic
IVCs and HVCs has unveiled the presence of sub-pc/AU scale gaseous
structures in the Milky Way's extra-planar ISM. 
Most of the H$_2$ absorption in the IVCs in the lower 
Galactic halo appears to be associated with the
cold neutral medium (CNM) at temperatures of $T\leq 300$ K, 
volume densities of $n_{\rm H}\approx 30$ cm$^{-3}$, and linear
diameters of $D\approx 0.1$ pc. The CNM in IVCs 
has a relatively large area filling factor and thus
must represent a gas phase that is characteristic
for the denser, neutral regions in the halo. The
detection of H$_2$ absorption in an IVC towards the
LMC star Sk $-$68\,80 has shown that even smaller
filaments at AU scale and with very high densities 
(almost $10^3$ cm$^{-3}$) exist in the halo, possibly
representing tiny-scale atomic structures (TSAS). This
finding implies that the Milky Way's extra-planar gas
consists of extreme small-scale structure. Many aspects
that concern the physical nature of these filaments 
(e.g., formation processes, thermal pressures, 
dust content, etc.) are not well understood yet, and more data
is required to explore these intriguing objects in more detail.

\acknowledgements{P. Richter is supported by the 
{\it Deutsche Forschungsgemeinschaft} through Emmy-Noether Fellowship RI 1124/3-1.}


\begin{thebibliography}{}

\bibitem[]{}
Bluhm, H., de\,Boer, K.S., Marggraf, O., \& Richter, P. 2001, A\&A 367, 299

\bibitem[]{}
Faison, M.D., Goss, W.M., Diamond, P.J., \& Taylor, G.B. 1998, AJ 116, 2916

\bibitem[]{} 
Gringel, W., Barnstedt, J., de Boer, K.S., Grewing, M., Kappelmann, N., 
Richter, P. 2000, A\&A, 358, L38

\bibitem[]{}
Heiles, C. 1997, ApJ 481, 193

\bibitem[]{}
Lu, L., Savage, B.D., Sembach, K.R., Wakker, B.P., Sargent, W.L.W., \&
Osterloo, T.A. 1998, AJ, 115, 162 

\bibitem[]{}
Meyer, D.M., \& Lauroesch, J.T. 1999, ApJ 520, L103

\bibitem[]{}
Moos, H.W., et al. 2000, ApJ 538, L1

\bibitem[]{}
Richter, P., de Boer, K.S., Widmann, H., et al. 1999, Nature, 402, 386

\bibitem[]{}
Richter, P., Savage, B.D., Wakker, B.P., Sembach, \& Kalberla, P.M.W.
2001a, ApJ, 549, 281

\bibitem[]{}
Richter, P., Savage, B.D., Wakker, B.P., Sembach, K.R., Tripp, T.M.,
Murphy, E.M., Kalberla, P.M.W., \& Jenkins, E.B. 2001b, ApJ, 559, 318

\bibitem[]{}
Richter, P., Sembach, K.R., Wakker, B.P., \& Savage, B.D. 2001c, 
ApJ, 562, L181

\bibitem[]{}
Richter, P., Wakker, B.P., Savage, B.D., \& Sembach, K.R. 2003,
ApJ, 586, 230

\bibitem[]{}
Richter, P., Sembach, K.R., \& Howk, J.C. 2003, A\&A, 405, 1013

\bibitem[]{} 
Savage, B.D., Bohlin, R.C., Drake, J.F., Budich, W. 1977, ApJ, 216, 291

\bibitem[]{} 
Sembach, K.R., Howk, J.C., Savage, B.D., \& Shull, J.M. 2001, AJ, 121, 992

\bibitem[]{}
Shapiro, P.R., \& Field, G.B. 1976, ApJ, 205, 762

\bibitem[]{}
Spitzer, L. 1978, `Physical Processes in the Interstellar Medium', 
Wileys Classics Library, ISBN 0-471-02232-2

\bibitem[]{}
Wakker, B.P., Howk, J.C., Savage, B.D., et al. 1999, Nature, 402, 388

\bibitem[]{}
Wolfire, M.G., McKee, C.F., Hollenbach, D., Tielens, A.G.G.M. 1995, ApJ, 453, 673 

\end{thebibliography}
\end{document}